\documentclass[10pt,conference]{IEEEtran}

\usepackage{cite}
\usepackage{amsmath,amssymb,amsfonts}
\usepackage{amsthm}
\usepackage{graphicx}
\usepackage{textcomp}
\usepackage{xcolor}
\usepackage{algorithm}
\usepackage{algpseudocode}
\usepackage{booktabs}
\usepackage{multirow}
\usepackage[caption=false,font=footnotesize]{subfig}
\usepackage{array}
\usepackage{url}
\usepackage{tikz}
\usepackage{quantikz}
\usetikzlibrary{arrows.meta, positioning, shapes.geometric, fit, backgrounds}
\usepackage[hyphens]{xurl}
\usepackage{microtype}
\usepackage{balance}
\usepackage[hidelinks,breaklinks=true]{hyperref}
\usepackage{orcidlink}

\hypersetup{
  pdftitle={ConQEC: Concolic Testing for Quantum Error-Correcting Code Implementations},
  pdfauthor={Zihao Li, Ji Guan, Mingsheng Ying},
  pdfkeywords={quantum software testing, quantum error correction, concolic testing, symbolic execution, automated test generation, debugging, failure witnesses}
}

\graphicspath{{figures/}}
\newcolumntype{L}[1]{>{\raggedright\arraybackslash}p{#1}}

\def\BibTeX{{\rm B\kern-.05em{\sc i\kern-.025em b}\kern-.08em
    T\kern-.1667em\lower.7ex\hbox{E}\kern-.125emX}}

\theoremstyle{definition}

\begin{document}
\bstctlcite{IEEEtran:BSTcontrol}

\title{ConQEC: Concolic Testing for\\Quantum Error-Correcting Code Implementations}

\author{
\IEEEauthorblockN{
Zihao Li\textsuperscript{1,2,3}~\orcidlink{0009-0002-1768-6662},
Ji Guan\textsuperscript{1,2,*}~\orcidlink{0000-0002-3490-0029},
Mingsheng Ying\textsuperscript{4}~\orcidlink{0000-0003-4847-702X}
}
\IEEEauthorblockA{
\textsuperscript{1}\textit{Key Laboratory of System Software
(Chinese Academy of Sciences)}, Beijing, China
}
\IEEEauthorblockA{
\textsuperscript{2}\textit{Institute of Software,
Chinese Academy of Sciences}, Beijing, China
}
\IEEEauthorblockA{
\textsuperscript{3}\textit{University of Chinese Academy of Sciences},
Beijing, China
}
\IEEEauthorblockA{
\textsuperscript{4}\textit{Centre for Quantum Software and Information,
University of Technology Sydney}, Sydney, Australia
}
\IEEEauthorblockA{
lizh@ios.ac.cn;
guanj@ios.ac.cn;
Mingsheng.Ying@uts.edu.au
}
}

\maketitle
\pagestyle{plain}
\thispagestyle{plain}

\begingroup
\renewcommand{\thefootnote}{*}
\footnotetext{Corresponding author.}
\endgroup

\begin{abstract}
Quantum error-correcting code (QECC) implementations are hybrid software-control systems: measured syndromes drive decoder dispatch, recovery actions, and logical checks. Existing QECC verification techniques can prove formal obligations, and general quantum testing can expose final failures, but developers still lack a scalable testing workflow that generates concrete faults and explains where a QECC repair path first diverges. This paper presents ConQEC, to our knowledge the first concolic testing framework for QECC implementations. By combining concrete (Pauli) fault replay with symbolic repair-path constraints, ConQEC generates tests that expose not only final logical failures but also fault--syndrome--decoder--recovery (FSDR) diagnostics: the injected fault, observed syndrome, decoder decision, recovery action, oracle result, and earliest divergent stage. Its symbolic encoding of Pauli faults, syndrome predicates, decoder and recovery relations, and oracle checks enables this diagnostic workflow to scale beyond exponentially increasing amplitude-level quantum-state reasoning. We implemented ConQEC and evaluated it on generated source-level QECC families and public QECC artifacts with up to thousands of qubits. ConQEC detects the seeded bug in every experimental case and reports a replay-valid, fault-localizing FSDR witness for each.

\end{abstract}


\section{Introduction}
\label{sec:intro}
Quantum programs are increasingly becoming components of larger software stacks. Recent work studies coverage criteria and coverage-guided test
generation~\cite{ali2021coverage,wang2021quito,ye2023quratest},
bug studies and bug benchmarks~\cite{paltenghi2022bugs,zhao2023bugs4q},
static analysis~\cite{qchecker2023arxiv}, runtime
assertions~\cite{huang2019statistical,li2020projection}, and compiler
verification~\cite{tao2022giallar}.
Together, these efforts show that quantum software increasingly raises familiar software-engineering questions: how to generate effective tests, measure meaningful coverage, and localize implementation failures under quantum-specific semantics.

Quantum error-correcting code (QECC) implementations make this need especially
acute. QECCs are the basic mechanism by which fault-tolerant quantum computers
protect logical quantum information from noisy physical
hardware~\cite{shor1995scheme,steane1996error,calderbank1996good,gottesman1997stabilizer,dennis2002topological,fowler2012surface,terhal2015quantum}.
Instead of storing information in a single fragile physical qubit (quantum bit),
a QECC encodes it into a larger block of physical qubits, repeatedly extracts syndrome
information to infer possible faults, and applies recovery operations intended
to restore the logical state. For the fault inputs to this pipeline, we focus on
Pauli faults because they are the standard discrete abstraction used by QECC
decoders and because general quantum errors can be decomposed into Pauli
components~\cite{nielsen2010quantum}. In practice, this process is implemented
as a hybrid software-control pipeline: syndrome extraction produces classical
bits, decoder software maps those bits to repair decisions, recovery code
applies conditional corrections, and logical checks observe whether the repair
succeeded. This creates an implementation gap between the mathematical code and
the executable QECC stack. A code theorem may be correct while the
implementation reads the wrong measurement bit, dispatches to the wrong decoder
arm, omits a recovery case, or applies a correction at the wrong location. The
result may be a final logical failure, but the developer still needs to know
where the repair path first diverged.

This makes QECC bug finding a software-engineering problem. A useful QECC bug
report is not just a failed final assertion; it is a replayable repair-path
witness: the concrete fault, observed syndrome, decoder decision, recovery
action, oracle result, and earliest divergent stage. Such a witness can be
inspected, minimized, and stored as a regression test
\cite{zeller2002simplifying}. Without this staged evidence, a QECC testing tool
may reveal that something failed while leaving the software bug difficult to
localize.

Most existing work on QECC correctness has been driven by formal methods for
verification. Formal-verification tools and frameworks such as
QSE~\cite{fang2024qse}, QECV~\cite{wu2021qecv},
Veri-QEC~\cite{huang2025veriqec}, and fault-tolerance
verification~\cite{chen2025faulttolerance} have shown how to symbolically
execute, check, or verify QEC/QECC programs against formal correctness
obligations. These techniques are important foundations, but they do not
directly provide the testing workflow above: their primary goal is verification
rather than budgeted test generation and developer-facing debugging. They do not
directly answer which concrete Pauli fault should be generated next, which
syndrome/decoder/recovery behaviors have been exercised, or how a failure should
be returned as a replayable debugging witness.

Classical concolic testing suggests a natural direction because QECC
implementations are branch-heavy programs. Syndrome bits determine decoder
branches; decoder decisions determine recovery actions; and logical checks
expose whether the repair succeeded. Classical concolic testing systematically
steers inputs by recording path constraints and solving modified branch
conditions~\cite{godefroid2005dart,sen2005cute,cadar2008klee,cadar2013symbolic}.
However, a direct transfer to QECC is not sufficient. QECC inputs are large
discrete Pauli fault spaces, intermediate executions are staged repair paths
over syndromes, decoder decisions, and recovery actions, and useful coverage is
not ordinary source-branch coverage but QECC-specific semantic coverage.
Quantum Concolic Testing (QCT)~\cite{xia2025qct}, a concolic-testing framework
for general quantum programs, takes an important first step in this direction.
However, its amplitude-based constraint model is difficult to scale to large
QECC implementations because amplitude-level reasoning must track a quantum
state space that grows exponentially with the number of physical qubits; its
reported branch-coverage evaluation focuses on programs with fewer than five
qubits. This scale gap is no longer hypothetical: quantum hardware has already
crossed the thousand-physical-qubit threshold, as illustrated by IBM's
1,121-qubit Condor processor~\cite{ibm2023condor}, and practical QECC software
must manage large numbers of physical qubits, checks, syndromes, and recovery
decisions. Existing formal-verification and general quantum concolic-testing
approaches therefore do not yet provide a scalable, diagnostic testing workflow
for such QECC implementations. This leaves the central problem of this paper:

\begin{quote}
\itshape
How can concolic testing be made scalable to thousand-qubit QECC implementations
while still producing diagnostically useful failure witnesses?
\end{quote}

To answer this question, we propose ConQEC, a concolic-testing framework for
QECC implementations. ConQEC addresses the scalability and diagnostic challenges
with two design ideas. First, it treats each test as a Fault--Syndrome--Decoder--Recovery (FSDR) repair path and accepts a bug only 
after actual/reference replay produces a staged witness.
This makes QECC bug
reports developer-facing: a failure is not only a final oracle violation, but a replayable explanation of the first repair-path stage where the
implementation diverges from the reference. Second,
ConQEC makes test generation symbolic at the repair-path level: Pauli faults are
represented as symbolic variables and propagated through syndrome, detector,
stabilizer, decoder, recovery, and oracle predicates. This compact
representation lets ConQEC solve for targeted repair-path behaviors instead of
enumerating concrete faults, enabling concolic testing to scale to large QECC
implementations while preserving replay evidence.

\paragraph*{\textbf{Summary of Our Contributions}} This paper contributes:
\begin{itemize}
\item \textbf{ConQEC.}
We present, to the best of our knowledge, the first concolic testing framework
for QECC implementations. ConQEC models tests as FSDR repair paths and returns replay-backed
witnesses that localize the earliest actual/reference divergence.

\item \textbf{Scalable repair-path generation.}
We encode Pauli faults, syndromes, decoder/recovery relations, and oracle checks
as repair-path constraints, avoiding amplitude-level state reasoning while
targeting thousand-qubit QECC behaviors.

\item \textbf{Evaluation evidence.} We evaluate ConQEC on generated source-level QECC families and public QECC
artifacts with up to thousands of qubits. ConQEC detects the seeded bug in every
experimental case and reports a replay-valid, fault-localizing FSDR witness for
each, while also providing semantic-coverage and scale evidence beyond toy QECC
instances.
\end{itemize}

\section{Background and Motivating Example}
\label{sec:background}
\noindent This section gives the minimum background needed to follow ConQEC. We
first summarize the syndrome-based QECC repair pipeline and the concolic testing
loop, then use a small QECC bug to show why a useful test report should be a
replayable FSDR witness rather than only a
final pass/fail result.

\subsection{Quantum Error-Correcting Code}
\label{sec:qecc-background}

A QECC protects logical quantum information by spreading it across multiple
physical qubits (quantum bits), so that local physical faults can be detected
and corrected without directly measuring the protected logical state
\cite{shor1995scheme,steane1996error,calderbank1996good,gottesman1997stabilizer,dennis2002topological,fowler2012surface,terhal2015quantum}.
The common software pattern is syndrome-based repair: parity-check measurements
produce classical syndrome bits indicating which checks are violated; a decoder
maps the syndrome to a correction; and recovery applies that correction to
restore the intended logical behavior. We call this staged sequence a repair
path. Figure~\ref{fig:qecc-background-flow} summarizes the process. 

The motivating example below instantiates this repair path in a small executable
QECC implementation.

\definecolor{qecenc}{RGB}{225,238,255}
\definecolor{qecfault}{RGB}{255,232,232}
\definecolor{qecsyn}{RGB}{232,244,232}
\definecolor{qecdec}{RGB}{255,244,214}
\definecolor{qecrec}{RGB}{236,232,255}

\providecommand{\qeccstage}[6]{}
\renewcommand{\qeccstage}[6]{%
  \node[stagebox, fill=#3] (#1) at (#2,0) {};
  \node[stagetitle] at ([yshift=0.115in]#1.center) {#4};
  \node[stagesub]   at ([yshift=-0.020in]#1.center) {#5};
  \node[stagesub]   at ([yshift=-0.130in]#1.center) {#6};
}

\begin{figure}[h]
\centering
\resizebox{0.98\columnwidth}{!}{%
\begin{tikzpicture}[
  x=1in,
  y=1in,
  >=stealth,
  stagebox/.style={
    draw=black!55,
    line width=0.35pt,
    minimum width=0.64in,
    minimum height=0.45in,
    inner sep=0pt
  },
  stagetitle/.style={
    font=\fontsize{8.6pt}{8.6pt}\selectfont\bfseries,
    inner sep=0pt
  },
  stagesub/.style={
    font=\fontsize{7.25pt}{7.25pt}\selectfont,
    inner sep=0pt
  },
  flowarrow/.style={
    ->,
    line width=0.45pt,
    shorten >=1.5pt,
    shorten <=1.5pt
  }
]
\qeccstage{enc}{0.00}{qecenc}{Encode}{logical}{information}
\qeccstage{flt}{0.76}{qecfault}{Fault}{Pauli}{error}
\qeccstage{syn}{1.52}{qecsyn}{Syndrome}{parity}{check}
\qeccstage{dec}{2.28}{qecdec}{Decode}{choose}{correction}
\qeccstage{rec}{3.04}{qecrec}{Recover}{restore}{state}

\draw[flowarrow] (enc.east) -- (flt.west);
\draw[flowarrow] (flt.east) -- (syn.west);
\draw[flowarrow] (syn.east) -- (dec.west);
\draw[flowarrow] (dec.east) -- (rec.west);
\end{tikzpicture}}
\caption{Basic QECC correction process.}
\label{fig:qecc-background-flow}

\end{figure}
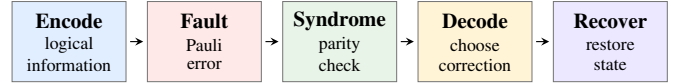

\subsection{Concolic Testing}
\label{sec:concolic-background}

Concolic testing, also known as dynamic symbolic execution, combines concrete
execution with symbolic constraint solving
\cite{godefroid2005dart,sen2005cute,cadar2008klee,cadar2013symbolic}: a tester
runs the program on one concrete input, records the path conditions encountered
during execution, flips one condition, and asks a solver for a new input that
exercises a different path. Repeating this loop steers testing through program
behavior rather than relying only on random inputs. In QECC implementations,
the concrete input is a Pauli fault, and the path of interest is the repair
path from syndrome extraction through decoder choice and recovery action to the
final logical check.

\subsection{Motivating Example}
\label{sec:motivating}

We use a three-qubit repetition  code~\cite{nielsen2010quantum} because it is the smallest example that
still contains the repair stages ConQEC targets. For bit-flip faults, it is
analogous to a classical three-copy code: information is spread across
three physical qubits labeled \(q_0\), \(q_1\), and \(q_2\), and a single
flipped position can be inferred by comparing neighboring positions. The quantum
version performs these comparisons as syndrome measurements, which reveal
violated checks rather than directly reading the protected logical state. 

\makeatletter
\@ifundefined{textcolor}{%
  \providecommand{\codekw}[1]{\textbf{\ttfamily #1}}%
  \providecommand{\codebug}[1]{\fbox{\textbf{\ttfamily #1}}}%
  \providecommand{\codett}[1]{{\ttfamily #1}}%
  \providecommand{\codebad}[1]{{\ttfamily #1}}%
  \providecommand{\codestage}[1]{\textbf{\ttfamily #1}}%
}{%
  \providecommand{\codekw}[1]{\textcolor{blue}{\textbf{\ttfamily #1}}}%
  \providecommand{\codebug}[1]{\fcolorbox{red}{white}{\textcolor{red}{\textbf{\ttfamily #1}}}}%
  \providecommand{\codett}[1]{{\ttfamily #1}}%
  \providecommand{\codebad}[1]{\textcolor{red}{\ttfamily #1}}%
  \providecommand{\codestage}[1]{\textcolor{green!40!black}{\textbf{\ttfamily #1}}}%
}
\makeatother

\begin{figure}[t]
\centering
\scriptsize
\setlength{\tabcolsep}{2pt}
\renewcommand{\arraystretch}{1.02}
\begin{tabular}{@{}p{0.46\columnwidth}|p{0.46\columnwidth}@{}}
\multicolumn{1}{c|}{\textbf{Reference implementation}} &
\multicolumn{1}{c}{\textbf{Buggy implementation}}\\
\hline
\multicolumn{2}{l}{\codestage{// Encode: logical information into physical qubits}}\\[-1pt]
{\codekw{encodeRep3}\codett{(q0,q1,q2)}} &
{\codekw{encodeRep3}\codett{(q0,q1,q2)}}\\[0.1em]

\multicolumn{2}{l}{\codestage{// Fault: injected Pauli error X@q0}}\\[-1pt]
{\codekw{inject}\codett{(X@q0)}} &
{\codekw{inject}\codett{(X@q0)}}\\[0.1em]

\multicolumn{2}{l}{\codestage{// Syndrome: parity-check measurements}}\\[-1pt]
{\codett{s0 = }\codekw{measureZZ}\codett{(q0,q1)}} &
{\codett{s0 = }\codekw{measureZZ}\codett{(q0,q1)}}\\
{\codett{s1 = }\codekw{measureZZ}\codett{(q1,q2)}} &
{\codett{s1 = }\codekw{measureZZ}\codett{(q1,q2)}}\\
{\codett{s = (s0,s1)}} &
{\codett{s = (s0,s1)}}\\[0.2em]

\multicolumn{2}{l}{\codestage{// Decode: choose correction from syndrome}}\\[-1pt]
{\codekw{if}\codett{ s == (1,0):}} &
{\codekw{if}\codett{ s == (1,0):}}\\
{\codett{\hspace*{0.8em}d = X@q0}} &
{\codett{\hspace*{0.8em}d = }\codebad{X@q1}\codett{ }\codebug{BUG}}\\
{\codekw{elif}\codett{ s == (1,1):}} &
{\codekw{elif}\codett{ s == (1,1):}}\\
{\codett{\hspace*{0.8em}d = X@q1}} &
{\codett{\hspace*{0.8em}d = X@q1}}\\
{\codekw{elif}\codett{ s == (0,1):}} &
{\codekw{elif}\codett{ s == (0,1):}}\\
{\codett{\hspace*{0.8em}d = X@q2}} &
{\codett{\hspace*{0.8em}d = X@q2}}\\
{\codekw{else}\codett{: d = I}} &
{\codekw{else}\codett{: d = I}}\\[0.2em]

\multicolumn{2}{l}{\codestage{// Recover: apply correction and check state}}\\[-1pt]
{\codekw{apply}\codett{(d); }\codekw{logicalCheck}\codett{()}} &
{\codekw{apply}\codett{(d); }\codekw{logicalCheck}\codett{()}}\\
\end{tabular}

\caption{Repair-path bug in the three-bit repetition code. For fault
\(X@q_0\), both programs reach syndrome \((1,0)\); the first divergence is
decoder, where the reference selects \(X@q_0\) and the buggy
implementation selects \(X@q_1\).}
\label{fig:motivating-repair-path-code}
\end{figure}

Figure~\ref{fig:motivating-repair-path-code} shows the example as a small
syndrome schematic and paired pseudocode. Reading Figure~\ref{fig:motivating-repair-path-code} from top to bottom,
\texttt{encodeRep3} is the encoding stage. We write \(P@q_i\) for applying
Pauli operator \(P\) at physical qubit \(q_i\). The Pauli-\(X\) operator is the
quantum analogue of a NOT gate: the kets \(|0\rangle\) and \(|1\rangle\) denote
the two standard basis states of a qubit, corresponding to classical 0 and 1,
and \(X\) swaps them. Thus \(X@q_0\) in the fault stage denotes a bit-flip fault
on \(q_0\).

A syndrome is the pattern of check outcomes used to diagnose the fault. Here,
each syndrome bit says whether one neighboring-qubit parity check is violated;
the pair \((s_0,s_1)\) is analogous to check bits in a classical error-correcting
code. In the pseudocode, this pair is stored as \texttt{s = (s0,s1)}; in
formulas, we write the same syndrome as \(\sigma\). A call such as
\texttt{measureZZ(q0,q1)} returns one such syndrome bit by comparing the
neighboring qubits \(q_0\) and \(q_1\). The \texttt{if}/\texttt{elif} block is
the decoder: \texttt{d} stores its chosen correction, \(I\) denotes no
correction, and \texttt{apply(d)} performs recovery.

For fault \(X@q_0\), only the first neighbor comparison is violated, so the code
variable \texttt{s}, written as \(\sigma\) in formulas, is \((1,0)\). The bug is
therefore a decoder error: the reference chooses \(X@q_0\), while the
implementation chooses \(X@q_1\) for the same reachable syndrome.

This example illustrates why final failures are not enough. A
\texttt{logicalCheck()} failure says that recovery did not work, but it does not
say whether the failure came from an incorrect syndrome, decoder branch,
recovery action, or final check. A useful report should
instead replay the repair path and identify the first divergent stage.

For this example, the ConQEC workflow introduced next would generate fault
\(X@q_0\), replay it against both the implementation and the reference contract,
and report an FSDR witness for the decoder branch at \(\sigma=(1,0)\): the
implementation chooses \(X@q_1\), while the reference chooses \(X@q_0\).

\section{ConQEC Design}
\label{sec:approach}
\begin{figure*}[t]
  \centering
  \includegraphics[width=0.8\textwidth]{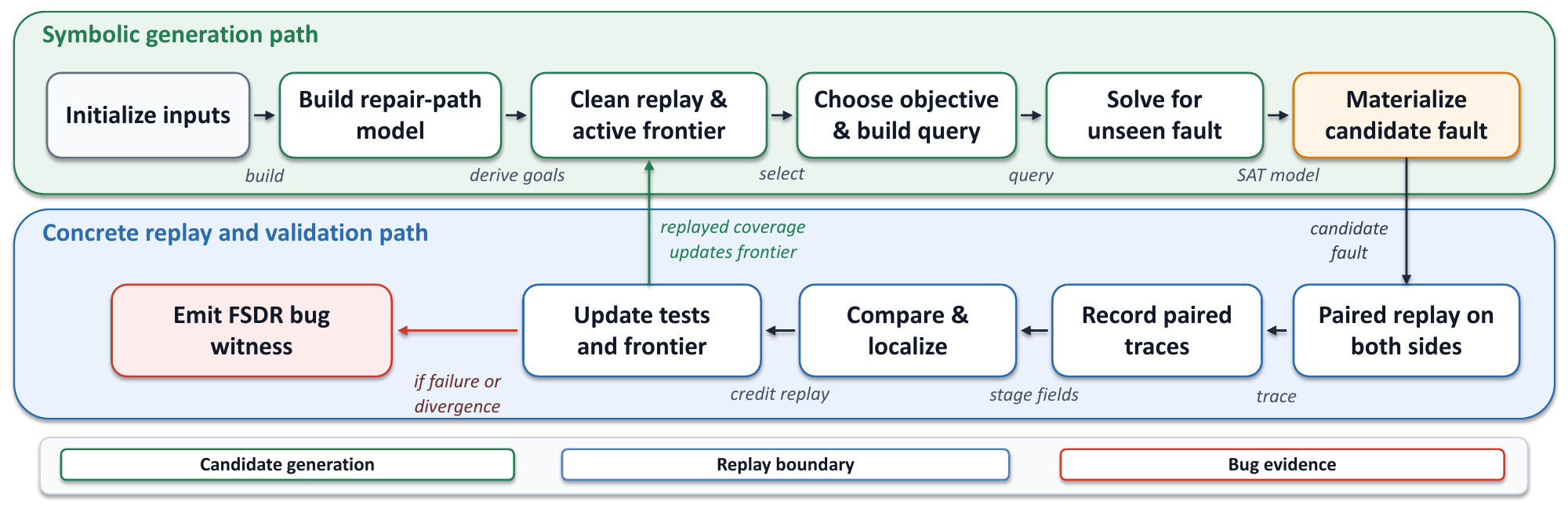}
  \caption{ConQEC workflow: symbolic repair-path constraints propose Pauli
  faults, and paired replay turns them into semantic coverage and
  FSDR bug witnesses.}
  \label{fig:conqec-workflow}
\end{figure*}

Section~\ref{sec:motivating} showed the testing problem through a concrete
QECC implementation: a final logical failure is insufficient unless the test
also explains which repair-path stage first diverged. ConQEC addresses this by
separating test generation from bug reporting: a symbolic model proposes
candidate Pauli faults, while paired replay against a reference contract decides
coverage credit and accepts replay-backed debugging witnesses. This section
presents the design as one workflow. We first define the replay-backed evidence
contract and the repair-path bug classes that ConQEC can localize. We then
introduce the symbolic repair-path model used to represent many Pauli-fault
executions compactly. Next, we describe the semantic objectives and query
interface that turn this model into candidate tests. Finally, we present the
concolic loop with \emph{paired replay}, meaning replaying the same concrete
fault on the implementation and the reference contract.
Figure~\ref{fig:conqec-workflow} summarizes this workflow.

\subsection{Evidence Contract and Bug Classes}
\label{sec:evidence-contract}

The evidence contract defines what ConQEC is allowed to report. A solver-produced
fault is only a candidate; it becomes testing evidence after replay, and it
becomes a bug witness only after the same fault is compared with the reference
contract. The reference contract records the expected repair-path behavior, such
as the expected syndrome, decoder decision, recovery action, or final logical
check. ConQEC therefore treats a test as a staged repair-path execution rather
than as a final pass/fail outcome. A concrete test is a replayed QECC execution
under the declared fault model:
\[
  \tau=\langle F,\bar\sigma_{\mathrm{impl}},d_{\mathrm{impl}},
  r_{\mathrm{impl}},\omega\rangle .
\]
Here \(F\) is the injected Pauli fault. The term
\(\bar\sigma_{\mathrm{impl}}\) denotes the implementation-side syndrome
evidence exposed during replay, including the produced syndrome
\(\sigma_{\mathrm{prod}}\) and, when observable, the syndrome value consumed by
later software \(\sigma_{\mathrm{use}}\). The term \(d_{\mathrm{impl}}\) is the
implementation's decoder decision, \(r_{\mathrm{impl}}\) is the recovery action
or normalized recovery result, and \(\omega\in\{\mathrm{PASS},\mathrm{FAIL}\}\) is
the replay-side oracle result, i.e., whether the final logical check or declared
postcondition succeeds. This trace is used for objective selection and coverage
updates.

A replayed candidate is promoted to an FSDR bug witness only after the same
fault is compared with the reference contract:
\[
  w_{\mathrm{bug}}=\langle F,\bar\sigma,d_{\mathrm{impl}},d_{\mathrm{ref}},
  r_{\mathrm{impl}},r_{\mathrm{ref}},\omega,\delta\rangle .
\]
The reference fields come from the declared reference behavior. The term
\(\bar\sigma\) denotes the replay-exposed syndrome evidence used for
classification:
\[
  \bar\sigma =
  \langle \sigma_{\mathrm{prod}},
  \sigma_{\mathrm{use}},
  \sigma_{\mathrm{ref}}\rangle ,
\]
where \(\sigma_{\mathrm{ref}}\) is the reference syndrome when it is available.
The term \(\delta\) is the earliest divergent repair-path stage. We use
\(\delta=\bot\) when replay finds no stage-level disagreement.
The same bottom symbol denotes absence in context: \(d_{\mathrm{impl}}=\bot\)
means that the implementation made no decoder decision for a reachable syndrome,
whereas \(\delta=\bot\) means that replay found no divergent stage.
If a field is not exposed by a subject, ConQEC skips bug-class rules that depend
on that field and still records the observable parts of the witness.

The bug class of a witness is defined by the earliest exposed repair-path stage
at which the implementation can be distinguished from the reference. ConQEC
applies the following stage-ordered rules:

\begin{itemize}
\item Syndrome extraction error:
\[
  \sigma_{\mathrm{prod}}\ne\sigma_{\mathrm{ref}} .
\]

\item Measurement register misuse:
\[
  (\sigma_{\mathrm{prod}}=\sigma_{\mathrm{ref}})
  \land
  (\sigma_{\mathrm{use}}\ne\sigma_{\mathrm{prod}}).
\]

\item Decoder misdispatch:
\[
  \sigma_{\mathrm{use}}=\sigma_{\mathrm{ref}}
  \land d_{\mathrm{impl}}\ne\bot
  \land d_{\mathrm{impl}}\ne d_{\mathrm{ref}} .
\]

\item Unhandled syndrome:
\[
  \sigma_{\mathrm{use}}=\sigma_{\mathrm{ref}}
  \land d_{\mathrm{impl}}=\bot
  \land d_{\mathrm{ref}}\ne\bot .
\]

\item Recovery contract violation:
\[
  d_{\mathrm{impl}}=d_{\mathrm{ref}}
  \land
  (r_{\mathrm{impl}}\ne r_{\mathrm{ref}}
  \lor \omega=\mathrm{FAIL}) .
\]
\end{itemize}
The rules are evaluated in order, so each witness is assigned to the earliest
applicable repair-path stage. The equations define the classification boundary;
Table~\ref{tab:repair-path-bug-classes} gives the developer-facing
interpretation.

\begin{table}[t]
\centering
\caption{ConQEC repair-path bug classes.}
\label{tab:repair-path-bug-classes}
\setlength{\tabcolsep}{3pt}
\renewcommand{\arraystretch}{1.12}
\begin{tabular}{@{}L{0.34\columnwidth}L{0.60\columnwidth}@{}}
\hline
\textbf{Class} & \textbf{Interpretation for developers} \\
\hline
\begin{tabular}[t]{@{}l@{}}Syndrome extraction\\error\end{tabular} &
The implementation produces a wrong check or detector value. \\

\begin{tabular}[t]{@{}l@{}}Measurement register\\misuse\end{tabular} &
The syndrome is produced correctly, but later software reads a stale, swapped,
or wrong measurement register. \\

\begin{tabular}[t]{@{}l@{}}Decoder misdispatch\end{tabular} &
A reachable syndrome is dispatched to the wrong decoder arm. \\

\begin{tabular}[t]{@{}l@{}}Unhandled syndrome\end{tabular} &
A reachable syndrome has no implementation handler. \\

\begin{tabular}[t]{@{}l@{}}Recovery contract\\violation\end{tabular} &
The decoder agrees with the reference, but the recovery result or final
contract check fails. \\
\hline
\end{tabular}
\end{table}

\subsection{Symbolic Repair-Path Model}
\label{sec:symbolic-repair-path}

The symbolic repair-path model is the scalability step in the workflow. It is not
a full quantum state vector. Instead, it is a compact constraint model over the
objects used for test generation: injected Pauli faults, syndrome or detector
predicates, and available decoder and recovery behavior. The model generates
candidate faults; paired replay still provides the testing evidence.
For a software-testing reader, the fault indicators below are input bits, the
predicate expressions are semantic branch conditions, and the decoder/recovery
relations are dispatch behavior in the repair software.

Let the declared fault model be a finite set of fault atoms
\(\mathcal F=\{\alpha_i\mid i\in I\}\). Each atom \(\alpha_i\) represents one
legal Pauli injection site and Pauli effect. ConQEC assigns each atom a symbolic
Boolean indicator \(\hat f_i\), where \(\hat f_i=1\) means that \(\alpha_i\) is
selected in the candidate fault. The legal fault domain, including weight bounds
and allowed injection sites, is encoded by \(\Phi_{\mathcal F}(\hat f)\). For
example, in the three-qubit repetition-code example from
Figure~\ref{fig:motivating-repair-path-code}, we use \(I=\{0,1,2\}\) and
\(\alpha_i=(X@q_i)\). An at-most-one-\(X\)-fault campaign, including the clean
fault, is encoded as
\[
  \Phi_{\mathcal F}(\hat f)\equiv
  \hat f_0+\hat f_1+\hat f_2\le 1 .
\]
A solver assignment \(\hat f_0=1,\hat f_1=\hat f_2=0\) is therefore materialized
as the concrete fault
\[
  F=\{\alpha_0\}=\{X@q_0\}.
\]

For each exposed check-like predicate \(c\), such as a detector bit, stabilizer
check, or final logical check, ConQEC represents its value under symbolic faults
as an affine parity expression:
\[
  s_c(\hat f)=\beta_c\oplus
  \bigoplus_{i\in N(c)} \hat f_i .
\]
Here \(N(c)\subseteq I\) is the support of predicate \(c\): it contains exactly
the fault atoms whose selection flips \(c\). The offset
\(\beta_c\in\{0,1\}\) records the clean-run or normalized value of the predicate.
Intuitively, a check flips when an odd number of selected Pauli faults touch that
check.
The collection of these expressions induces the symbolic syndrome or detector
vector \(\sigma\). For the same example, the two parity checks
\texttt{measureZZ(q0,q1)} and \texttt{measureZZ(q1,q2)}, conventionally written
as \(Z_0Z_1\) and \(Z_1Z_2\), yield
\[
  \sigma_0=\hat f_0\oplus \hat f_1,
  \qquad
  \sigma_1=\hat f_1\oplus \hat f_2 .
\]
Thus ConQEC can ask for a fault that reaches the syndrome
\(\sigma=(1,0)\) without enumerating all concrete faults.

When decoder behavior is available to the test generator, ConQEC represents it
as a relation \(\mathcal D(\sigma,d)\). This relation means that syndrome or
detector vector \(\sigma\) can dispatch to decoder decision \(d\). The relation
may come from source-level decoder logic, a declared decoder table, or an
explicit reference contract. In the reference decoder from
Figure~\ref{fig:motivating-repair-path-code}, for example, the relation contains
tuples such as
\[
\begin{aligned}
  &\mathcal D_{\mathrm{ref}}((1,0),X@q_0),
  \mathcal D_{\mathrm{ref}}((1,1),X@q_1), \mathcal D_{\mathrm{ref}}((0,1),X@q_2).
\end{aligned}
\]
The buggy implementation instead contains
\[
  \mathcal D_{\mathrm{impl}}((1,0),X@q_1),
\]
which lets ConQEC formulate a decoder misdispatch objective over the same
reachable syndrome.
We use \(\mathcal D\) generically for decoder relations and write
\(\mathcal D_{\mathrm{impl}}\) and \(\mathcal D_{\mathrm{ref}}\) when the
implementation and reference relations must be distinguished.

Similarly, when recovery behavior is available, ConQEC represents it as a
relation \(\mathcal R(\sigma,d,r)\), meaning that syndrome \(\sigma\) and
decoder decision \(d\) lead to recovery result \(r\). In a simple Pauli-frame
implementation, \(r\) may be the applied
correction itself; for example, \(d=X@q_0\) leads to a recovery result
that applies \(X@q_0\). If decoder or recovery internals are not
available to the symbolic generator, ConQEC does not invent them; those values
are observed later during paired replay and can still appear in an FSDR witness.
The same convention applies to \(\mathcal R_{\mathrm{impl}}\) and
\(\mathcal R_{\mathrm{ref}}\).

The symbolic repair-path model is therefore
\[
  \mathcal S =
  \langle
  \hat f,\Phi_{\mathcal F},\{s_c\},\mathcal D,\mathcal R
  \rangle .
\]
Here \(\hat f\) and \(\Phi_{\mathcal F}\) describe the legal Pauli-fault space,
\(\{s_c\}\) describes how faults affect syndrome and detector predicates,
\(\mathcal D\) describes syndrome-to-decoder behavior when available, and
\(\mathcal R\) describes decoder-to-recovery behavior when available.

This representation is compact. A single symbolic repair-path model represents
many concrete Pauli fault patterns at once. ConQEC can therefore solve for
targeted repair-path behavior without enumerating the concrete fault space or
carrying an amplitude-level quantum state vector. Paired replay then turns the
selected symbolic candidate into concrete testing evidence.

\subsection{Objectives and Candidate Generation}
\label{sec:objectives-queries}

The symbolic model describes which repair-path behaviors are reachable; objectives
decide which behaviors the solver should try next. ConQEC uses coverage
objectives to target measurement outcomes, stabilizer conditions, and transition
sequences, and failure-directed objectives to target likely
implementation--reference divergences. Table~\ref{tab:objective-families}
summarizes the correspondence. The coverage denominator is therefore repair-path
behavior, not merely source branches. Solving an objective yields only a
candidate fault; paired replay decides whether it covers an obligation, exposes
a witness, or passes.

\begin{table}[t]
\centering
\caption{ConQEC objective layers and classical analogues.}
\label{tab:objective-families}
\setlength{\tabcolsep}{3pt}
\renewcommand{\arraystretch}{1.10}
\begin{tabular}{@{}
  >{\raggedright\arraybackslash}p{0.24\columnwidth}
  >{\raggedright\arraybackslash}p{0.24\columnwidth}
  >{\raggedright\arraybackslash}p{0.46\columnwidth}
@{}}
\hline
\textbf{QECC objective layer} &
\textbf{Classical analogue} &
\textbf{Interpretation for developers} \\
\hline

\(\mathcal O_B\): measurement outcome &
Branch coverage &
A check, detector, or final logical check reaches one of its Boolean outcomes. \\

\(\mathcal O_C\): stabilizer condition &
Condition coverage &
The local activation count distinguishes whether one or multiple faults affect
the same check. \\

\(\mathcal O_{TS}\): transition sequence &
Branch pair or path fragment coverage &
The repair path exercises an ordered pair leading to decoder or recovery behavior. \\

Failure-directed &
Bug-directed path objective &
The objective proposes a stage divergence that paired replay must validate. \\

\hline
\end{tabular}
\end{table}

For a concrete fault \(F\subseteq\mathcal F\), evaluating the symbolic indicator
\(\hat f_i\) under \(F\) gives
$
  \hat f_i(F)=1 \iff \alpha_i\in F .
$
Evaluating \(s_c(\hat f)\) under the same assignment gives the concrete
predicate value \(s_c(F)\). We write \(\sigma(F)\) for the syndrome vector
induced by the predicate values \(\{s_c(F)\}\).
In the objective definitions below, \(F\) ranges only over legal faults, meaning
its indicator valuation satisfies \(\Phi_{\mathcal F}\).

\textbf{Measurement-outcome objectives.}
For each exposed outcome predicate \(c\), such as a syndrome bit or detector
bit, ConQEC records its reachable Boolean outcomes:
\[
  \mathcal O_B
  =
  \{(c,b)\mid b\in\{0,1\},\ \exists F\subseteq\mathcal F.\ s_c(F)=b\}.
\]
A replayed test covers \((c,b)\) when the replayed fault \(F\) makes predicate
\(c\) take value \(b\). In the motivating example from
Figure~\ref{fig:motivating-repair-path-code}, \((\sigma_0,1)\) asks for
\(\sigma_0=\hat f_0\oplus\hat f_1=1\), which is satisfied by
\(F=\{X@q_0\}\).

\textbf{Stabilizer-condition objectives.}
Measurement outcomes alone do not distinguish how a predicate was reached. A
stabilizer-condition objective asks not only whether a check fired, but how many
selected fault atoms touched that check. ConQEC therefore records, for each
predicate \(c\), the number of selected fault atoms in its support. Symbolically,
this is
\[
  \mu_c(\hat f)=\sum_{i\in N(c)} \hat f_i .
\]
Under a concrete replayed fault \(F\), evaluating this expression gives
\(\mu_c(F)\).
The stabilizer-condition obligations are
\[
  \mathcal O_C
  =
  \{(c,k)\mid \exists F\subseteq\mathcal F.\ \mu_c(F)=k\}.
\]
In the same example, \(N(\sigma_0)=\{0,1\}\). Thus \(F=\{X@q_0\}\) covers
\(\mu_{\sigma_0}=1\). Under a weight-2 version of the same fault model,
\(F=\{X@q_0,X@q_1\}\) covers \(\mu_{\sigma_0}=2\), the pair-mask frontier used
to test whether multiple faults on the same check are distinguished.

\textbf{Transition-sequence objectives.}
These objectives record whether syndrome and stabilizer conditions lead to
specific decoder or recovery behavior. Using the decoder relation
\(\mathcal D\) and recovery relation \(\mathcal R\) from
Section~\ref{sec:symbolic-repair-path}, ConQEC defines
\[
\begin{aligned}
&\mathcal O_{TS} =
\{\,((c,k),d) \mid
    \exists F\subseteq\mathcal F.\ 
    \mu_c(F)=k \land \mathcal D(\sigma(F),d)\,\}\\
&{}\cup
\{\,(\sigma,d) \mid
    \exists F\subseteq\mathcal F.\ 
    \sigma(F)=\sigma \land \mathcal D(\sigma,d)\,\}\\
&{}\cup
\{\, (d,r) \mid
    \exists F\subseteq\mathcal F,\exists \sigma.\ 
    \sigma(F)=\sigma \land
    \mathcal D(\sigma,d) \land \mathcal R(\sigma,d,r)\,\}.
\end{aligned}
\]
These three groups link a local stabilizer condition to a decoder decision, a
syndrome value to a decoder decision, and a decoder decision to a recovery
result, respectively. Here \(\mathcal D\) and \(\mathcal R\) denote the
generator-visible relations used for coverage; objectives that compare the
implementation and reference write the \(\mathrm{impl}\) and \(\mathrm{ref}\)
variants explicitly. A replayed test covers a transition-sequence obligation
when its replayed repair path contains the corresponding ordered pair.

In the same example, \(F=\{X@q_0\}\) induces \(\sigma(F)=(1,0)\). The reference
path contains the transition
\[
  \sigma=(1,0)\rightarrow d=X@q_0,
\]
whereas the buggy implementation contains
\[
  \sigma=(1,0)\rightarrow d=X@q_1.
\]
Thus transition-sequence objectives capture not only which syndrome was reached,
but also which software action followed that syndrome.

\textbf{Failure-directed objectives.}
These objectives are not coverage obligations. They steer generation toward
likely repair-path divergences, such as
\[
\begin{array}{@{}l@{}}
D_{\mathrm{impl}}(\sigma)\ne D_{\mathrm{ref}}(\sigma),
D_{\mathrm{impl}}(\sigma)=\bot,
R_{\mathrm{impl}}(\sigma,d)\ne R_{\mathrm{ref}}(\sigma,d).
\end{array}
\]
Here \(D\) and \(R\) are functional views of the decoder and recovery behavior
represented by the relations \(\mathcal D\) and \(\mathcal R\) when those
internals are available to the generator.
For the motivating bug, the decoder-misdispatch objective asks for a reachable
\(\sigma\) such that \(D_{\mathrm{impl}}(\sigma)\ne D_{\mathrm{ref}}(\sigma)\);
a satisfying assignment may yield \(F=\{X@q_0\}\). It becomes a bug report only
after paired replay observes an oracle failure or an implementation--reference
stage divergence.

For a replayed test set \(T\), ConQEC reports each coverage layer with the same
denominator form:
\[
  \mathrm{Cov}_{L}(T)=
  \frac{|\mathcal O_L(T)|}{|\mathcal O_L|},
  \qquad
  L\in\{B,C,TS\}.
\]
Here \(\mathcal O_L(T)\) is the subset of obligations in \(\mathcal O_L\)
covered by at least one replayed fault in \(T\).

To materialize an objective \(o\), ConQEC builds a constraint query over
\(\mathcal S\). The query always includes the legal fault-domain constraint and
the syndrome equations; it then adds the constraint \(\Phi_o\) that represents
the selected objective:
\[
  q_o(\hat f,\sigma,d,r)
  =
  \Phi_{\mathcal F}(\hat f)
  \wedge
  \Phi_{\sigma}(\hat f,\sigma)
  \wedge
  \Phi_o(\hat f,\sigma,d,r).
\]
The formula \(\Phi_{\sigma}\) links fault variables to the symbolic syndrome
predicates, for example by enforcing \(\sigma_c=s_c(\hat f)\). The formula
\(\Phi_o\) is obtained directly from the selected objective \(o\). For a
measurement objective \(o=(c,b)\), \(\Phi_o\) is \(s_c(\hat f)=b\). For a
stabilizer-condition objective \(o=(c,k)\), it is
\(\mu_c(\hat f)=k\). For a transition-sequence objective,
\(\Phi_o\) constrains the decoder or recovery relation, such as
\(\mathcal D(\sigma,d)\) or \(\mathcal R(\sigma,d,r)\). For a failure-directed
objective, \(\Phi_o\) encodes a stage disagreement, such as
\(D_{\mathrm{impl}}(\sigma)\ne D_{\mathrm{ref}}(\sigma)\). A satisfying
assignment is materialized as a concrete Pauli fault \(F\), which is then
validated by paired replay.
Variables not used by a selected objective are existentially unconstrained and
ignored when materializing \(F\).

\subsection{Concolic Loop with Paired Replay}
\label{sec:loop-replay}

Algorithm~\ref{alg:conqec-loop} summarizes the ConQEC exploration loop. It
alternates between solver-guided candidate generation and replay-backed
validation: an SMT query selects a previously unseen fault for an uncovered
objective, and paired replay determines both coverage progress and bug-witness
acceptance.

The SMT line in Algorithm~\ref{alg:conqec-loop} uses the objective query
\(q_o\) from Section~\ref{sec:objectives-queries}, together with blocking
clauses that exclude concrete faults already recorded in \(T\). Each replay
record in \(T\) stores the tested fault and its paired replay traces, so an
``unseen'' fault means a fault not previously replayed. If the SMT query is
\(\textsc{Sat}\), the model gives a valuation \(\nu\) for the symbolic fault
indicators, which is materialized as
\[
  F=\{\alpha_i\in\mathcal F\mid \nu(\hat f_i)=1\}.
\]
The SMT model therefore proposes a concrete Pauli fault, but replay is still
required before ConQEC credits coverage or reports a bug.

The active frontier is replay-maintained. ConQEC initializes \(U\) from
repair-path-scoped semantic objectives extracted from the symbolic repair-path
model \(\mathcal S\). Paired replay determines which objectives are concretely
covered, and blocking clauses exclude concrete faults already recorded in \(T\).
Thus the solver proposes candidate faults for semantic objectives, while replay
is the only source of coverage credit and witness acceptance.

\begin{figure}[!t]
\begin{minipage}{\columnwidth}
\footnotesize
\hrule
\vspace{0.25ex}
\refstepcounter{algorithm}\label{alg:conqec-loop}
\noindent\textbf{Algorithm~\thealgorithm} ConQEC exploration loop
\vspace{0.25ex}
\hrule
\vspace{0.35ex}
\noindent\textbf{Input:} implementation \(C_{\mathrm{impl}}\), reference
\(C_{\mathrm{ref}}\), fault model \(\mathcal F\), budget \(B\).\\
\noindent\textbf{Output:} replayed tests \(T\), FSDR bug witnesses
\(W_{\mathrm{bug}}\)
\vspace{0.25ex}
\hrule
\vspace{0.3ex}
\begin{algorithmic}[1]
\Statex \textit{Step 1: Initialize}
\State build symbolic repair-path model \(\mathcal S\) from
\(C_{\mathrm{impl}}, C_{\mathrm{ref}}\), and \(\mathcal F\)
\State run paired replay on the clean fault \(F_0=\emptyset\)
\State record the clean replay, oracle result \(\omega_0\), and first divergence
\(\delta_0\)
\State initialize \(T\) with the clean replay record
\State initialize \(W_{\mathrm{bug}}\gets\emptyset\)
\If{\(\omega_0=\mathrm{FAIL}\) or \(\delta_0\ne\bot\)}
  \State construct and insert the clean FSDR bug witness into
  \(W_{\mathrm{bug}}\)
\EndIf
\State initialize \(U\) with coverage and failure-directed objectives

\While{\(B\) remains and \(U\neq\emptyset\)}
  \Statex \textit{Step 2: Generate a candidate fault}
  \State choose an uncovered objective \(o\in U\)
  \State solve the objective SMT query \(q_o\)
  from Section~\ref{sec:objectives-queries} for an unseen fault targeting \(o\)
  \If{the query is not \(\textsc{Sat}\)}
    \State remove \(o\) from \(U\)
    \State \textbf{continue}
  \EndIf
  \State materialize the concrete fault \(F\subseteq\mathcal F\) from the
  \(\textsc{Sat}\) model

  \Statex \textit{Step 3: Validate by paired replay}
  \State replay \(F\) on \(C_{\mathrm{impl}}\) and \(C_{\mathrm{ref}}\)
  \State record the implementation and reference traces
  \State compute the oracle result \(\omega\) and first divergent stage
  \(\delta\)

  \Statex \textit{Step 4: Update frontier and bug witnesses}
  \State add the replay record for \(F\) to \(T\)
  \State update covered objectives from this replay and remove them from \(U\)
  \If{\(\omega=\mathrm{FAIL}\) or \(\delta\ne\bot\)}
    \State construct the FSDR bug witness \(w_{\mathrm{bug}}\)
    \State \(W_{\mathrm{bug}}\gets W_{\mathrm{bug}}\cup\{w_{\mathrm{bug}}\}\)
  \EndIf
\EndWhile
\end{algorithmic}
\normalsize
\vspace{0.15ex}
\hrule
\end{minipage}
\end{figure}

The bug-witness acceptance rule uses a disjunction:
\[
  \omega=\mathrm{FAIL}\ \vee\ \delta\ne\bot
  \quad\Longrightarrow\quad
  \operatorname{AcceptBug}(w_{\mathrm{bug}}).
\]
The first case, \(\omega=\mathrm{FAIL}\), captures an end-to-end oracle failure.
The second case, \(\delta\ne\bot\), captures a contract-level repair-path
divergence between the implementation and the reference. It is not an arbitrary
trace difference: \(\delta\) records the first FSDR stage, such as syndrome,
decoder, or recovery behavior, where the paired replay violates the reference
contract. Thus ConQEC reports a bug witness either when the final oracle fails
or when replay exposes a source-visible repair-path bug even if the final oracle
is accidentally masked.

The accepted FSDR bug witness is
\[
  w_{\mathrm{bug}}=
  \langle F,\bar\sigma,d_{\mathrm{impl}},d_{\mathrm{ref}},
  r_{\mathrm{impl}},r_{\mathrm{ref}},\omega,\delta\rangle .
\]

For the motivating bug in Figure~\ref{fig:motivating-repair-path-code}, the
failure-directed objective asks for a reachable syndrome where the implementation
and reference decoders disagree. Under the at-most-one-\(X\)-fault domain, the
solver yields \(F=\{X@q_0\}\). Paired replay observes \(\sigma=(1,0)\),
\(d_{\mathrm{impl}}=X@q_1\), \(d_{\mathrm{ref}}=X@q_0\),
\(\omega=\mathrm{FAIL}\), and \(\delta=\mathrm{decoder}\). The candidate is
therefore accepted as an FSDR bug witness.

\section{Evaluation}
\label{sec:eval}
To evaluate ConQEC, we investigate the following four research questions:
\begin{itemize}
  \item \textbf{RQ1: Replay-backed fault localization.}
  Can ConQEC localize QECC repair-path faults as actionable
  FSDR witnesses at thousand-qubit scale?

  \item \textbf{RQ2: Semantic coverage analysis.}
  Does repair-path semantic guidance cover hard stabilizer
  obligations more effectively than strong valid-input random testing?

  \item \textbf{RQ3: Scale and cost.}
  How far does ConQEC scale on source-level QECC families, and
  which phase dominates the cost?

  \item \textbf{RQ4: Public artifact applicability.}
  Can ConQEC operate on recognized public QECC artifacts through
  supported adapters?
\end{itemize}

\subsection{Experimental Setup}
\label{sec:experimental-setup}

\textbf{Implementation.}
ConQEC is implemented as a testing layer over the QSE
symbolic-stabilizer backend~\cite{fang2024qse} and compatible adapter interfaces. For each
subject, the adapter exports the declared Pauli fault model, exposed
detector, stabilizer, or logical-observable predicates, optional decoder
or recovery relations, and paired replay entry points. ConQEC
builds semantic objectives from these exported objects, materializes
concrete Pauli faults from solver assignments, and credits coverage or
bug evidence only after replay.

\textbf{Subjects and artifacts.}
We use generated source-level QECC subjects drawn from standard QECC
families cataloged in the Error Correction Zoo~\cite{eczoo}, together
with public QECC artifacts. The source-level experiments use the
reported subject buckets: repetition, surface/toric, Tanner/LDPC-style
sparse-check subjects, one color-code control case, and fixed small
block-code subjects. The small-block bucket groups Steane, Shor, and
Hamming-CSS configurations rather than treating them as separate scale
families. RQ2 uses a controlled repetition-code ladder under legal
weight-2 \(X\)-faults, and RQ3 reports the source-scale frontier on
repetition, surface/toric, and Tanner/LDPC rows. The public-artifact benchmark is used in RQ4 and consists of the
supported non-\texttt{REPEAT} subset of the Tesseract public Stim
corpus~\cite{gidney2021stim,beni2025tesseract} and one PyMatching
historical parent/fixed regression~\cite{higgott2022pymatching}.

\textbf{Experimental environment.}
All experiments were run as isolated Docker jobs on an AMD EPYC 7702
server with 128 physical cores and 2.0~TiB memory, running
Ubuntu 24.04.4 LTS. The environment used Docker 29.4.3, Julia 1.10.10,
QuantumSE, and Z3 4.12.4.

\subsection{RQ1: Replay-Backed Fault Localization}

RQ1 evaluates whether ConQEC can localize seeded QECC
repair-path faults at thousand-qubit scale and turn them into
developer-usable FSDR witnesses. Applying 15 fixed source-level mutations to
all 13 configurations (Table~\ref{tab:rq1-family}) yields \(13\times15=195\)
mutants. Each mutant is replayed against its corresponding clean implementation
under the same fault to obtain the reference behavior. We instantiate three source-level mutation operators in each of the
five repair-stage bug classes:
\begin{itemize}
\item \textbf{Syndrome extraction error.}
The operators change the check-to-ancilla mapping, swap syndrome bits, or
drop/invert a check bit. For example, an extracted syndrome \(10\) may be
reported as \(01\).

\item \textbf{Measurement register misuse.}
The operators introduce a stale measurement read, a wrong-register read, or
an offset read. For example, the decoder may read the previous round's
syndrome register instead of the current round's register.

\item \textbf{Decoder misdispatch.}
The operators change the syndrome-to-arm mapping, retain a stale clean-arm
dispatch, or alias a neighboring arm. For example, the mapping
\(10\mapsto X@q_1\) may be changed to \(10\mapsto X@q_2\).

\item \textbf{Unhandled syndrome.}
The operators introduce a bottom-arm mapping, a no-op arm, or a dropped guard.
For example, the handler for the reachable syndrome \(10\) may be removed.

\item \textbf{Recovery contract violation.}
The operators change the Pauli correction, recovery target, or recovery
guard, leaving a residual error. For example, after correctly decoding
\(X@q_1\), the implementation may apply recovery to \(q_2\) instead of \(q_1\).
\end{itemize}

Operators are fixed
before generation and applied uniformly; generation receives no mutation
location, trigger, class, wrong target, or expected witness.

We measure localization
effort by a generation budget \(B\), consisting of one clean
seed and \(B-1\) non-clean attempts. A ConQEC attempt is a failure-objective
query; trace-order concolic (branch-order search) shares the symbolic machinery
but follows ordinary branch order without failure-directed guidance. Random
samples legal non-clean faults without replacement over 30 seeds. All methods
share the fault domain and replay criterion. Comparing the two concolic methods
isolates ConQEC's failure-directed guidance. A candidate fault is counted only after paired replay confirms an
FSDR record with the fault, syndrome evidence, decoder decision,
recovery behavior, oracle result, and earliest divergent stage.

Table~\ref{tab:rq1-budget} reports the budgeted results.
ConQEC localizes 156/195 seeded repair-path faults by \(B=2\),
and all 195 by \(B=4\). Under the same budget definition,
generic branch-order search finds no witnesses by \(B=4\), and
random fault testing finds only 15.7/195 witnesses on average
at \(B=4\). We also compare the methods under equal elapsed time budgets \(t\) per case,
using the same subjects, fault domain, and replay criterion. Random leads
slightly at 1\,s; ConQEC leads by 10\,s and reaches 195/195 by 60\,s.

\begin{table}[t]
\centering
\caption{RQ1 budgeted fault localization.}
\label{tab:rq1-budget}
\footnotesize
\begin{tabular}{@{}lccc@{}}
\toprule
Budget & ConQEC & Trace-order & Random \\
\midrule
\(B=2\)  & 156/195 & 0/195  & 4.6/195 \\
\(B=4\)  & 195/195 & 0/195  & 15.7/195 \\
\(B=8\)  & 195/195 & 5/195  & 36.6/195 \\
\(B=16\) & 195/195 & 66/195 & 68.8/195 \\
\midrule
\(t=1\,\mathrm{s}\)  & 105/195 & 0/195  & 110.4/195 \\
\(t=10\,\mathrm{s}\) & 150/195 & 60/195 & 142.2/195 \\
\(t=60\,\mathrm{s}\) & 195/195 & 75/195 & 175.4/195 \\
\bottomrule
\end{tabular}
\end{table}

Table~\ref{tab:rq1-class-effort} breaks down the replay-backed
witnesses (Wits.) by the five repair-path bug classes defined in
Section~\ref{sec:evidence-contract}. The stage labels denote syndrome extraction,
measurement use, decoder dispatch, and recovery; the two decoder-related
classes share the decoder dispatch stage. First witness budget \(B\)
counts the clean seed and attempts through the first accepted witness.
The first four classes yield witnesses at \(B=2\), on the first non-clean
query. Recovery contract violations first yield witnesses at \(B=4\), on the
third non-clean query, because the two earlier failure-directed objectives
are unsatisfiable.

\begin{table}[t]
\centering
\caption{RQ1 localization effort by bug class.}
\label{tab:rq1-class-effort}
\footnotesize
\setlength{\tabcolsep}{5pt}
\begin{tabular}{@{}lccl@{}}
\toprule
Bug class & Wits. & First witness & Divergence stage \\
\midrule
Syndrome extraction error & 39 & \(B=2\) & Syndrome extraction \\
Measurement register misuse & 39 & \(B=2\) & Measurement use \\
Decoder misdispatch & 39 & \(B=2\) & Decoder dispatch \\
Unhandled syndrome & 39 & \(B=2\) & Decoder dispatch \\
Recovery contract violation & 39 & \(B=4\) & Recovery \\
\midrule
All mutants & 195 & \(B=4\) & -- \\
\bottomrule
\end{tabular}
\end{table}

The accepted witnesses are not confined to the running example.
Table~\ref{tab:rq1-family} reports qubit and check counts, the number of source
configurations (Configs.), and witness counts across five QECC families.
The largest accepted
witness reaches 1500 physical qubits and 1499 checks; the largest non-repetition witness reaches 1058 qubits and 1058 checks; and
the largest sparse-check witness reaches 1000 qubits and 1000
checks.

\begin{table}[t]
\centering
\caption{RQ1 family-scale distribution.}
\label{tab:rq1-family}
\footnotesize
\begin{tabular}{@{}lcccc@{}}
\toprule
Family & Qubits & Checks & Configs. & Wits. \\
\midrule
Repetition & 1500 & 1499 & 3 & 45 \\
Surface/toric & 1058 & 1058 & 3 & 45 \\
Small block & 15 & 8 & 3 & 45 \\
Tanner/LDPC & 1000 & 1000 & 3 & 45 \\
Color & 7 & 6 & 1 & 15 \\
\midrule
Total & -- & -- & 13 & 195 \\
\bottomrule
\end{tabular}
\end{table}

Each accepted row is a replayed debugging artifact rather than
a solver-only candidate. The campaign therefore measures
replay-backed localization capability and generation effort for
seeded repair-path faults, not the natural frequency of these
bug classes.

\begin{center}
\fbox{%
\begin{minipage}{0.92\columnwidth}
\textbf{RQ1:}
ConQEC localizes seeded, thousand-qubit QECC repair-path faults
with substantially less generation effort than random fault
testing or generic branch-order search, while returning
replay-backed FSDR debugging witnesses.
\end{minipage}}
\end{center}

\subsection{RQ2: Semantic Coverage Analysis}

RQ2 evaluates whether repair-path semantic guidance covers hard stabilizer
obligations more effectively than strong valid-input random testing.
We compare both methods across the QECC counterparts of classical branch,
condition, and path-fragment coverage (Table~\ref{tab:objective-families}).

Measurement-outcome coverage records whether each exposed syndrome, detector,
or check predicate has been observed with both Boolean outcomes; it is the
closest analogue of branch-outcome coverage. Stabilizer-condition coverage
additionally distinguishes how many selected fault atoms lie in each check's
support; the declared fault model determines which counts are reachable
(Section~\ref{sec:objectives-queries}). For a check \(c\) with support \(N(c)\),
we define
\[
  \mu_c(F)=\sum_{i\in N(c)} f_i ,
\]
where \(f_i=1\) iff the fault atom at site \(i\) is selected in \(F\).

The transition-sequence layer gives
a path-fragment view of the repair path: when explicit \(D(\sigma,d)\) and
\(R(d,r)\) relations are available, it records ordered syndrome-to-decoder and
decoder-to-recovery transitions, credited only when the transition appears in
paired replay.

The criteria admit higher fault weights; here we choose legal weight-2
\(X\) faults as the simplest two-fault setting. We use a fixed-structure
repetition-code ladder on \(n=10,100,1000\) qubits, with check supports of size
two (\(d_c=2\)).
ConQEC and Random use the same domain of \(\binom{n}{2}\) legal fault pairs
and the same paired-replay acceptance rule. Both start from the same clean
seed, which is included in budget \(B\). Random samples pairs uniformly
without replacement.

\begin{figure*}[t]
\centering
\includegraphics[width=0.8\textwidth]{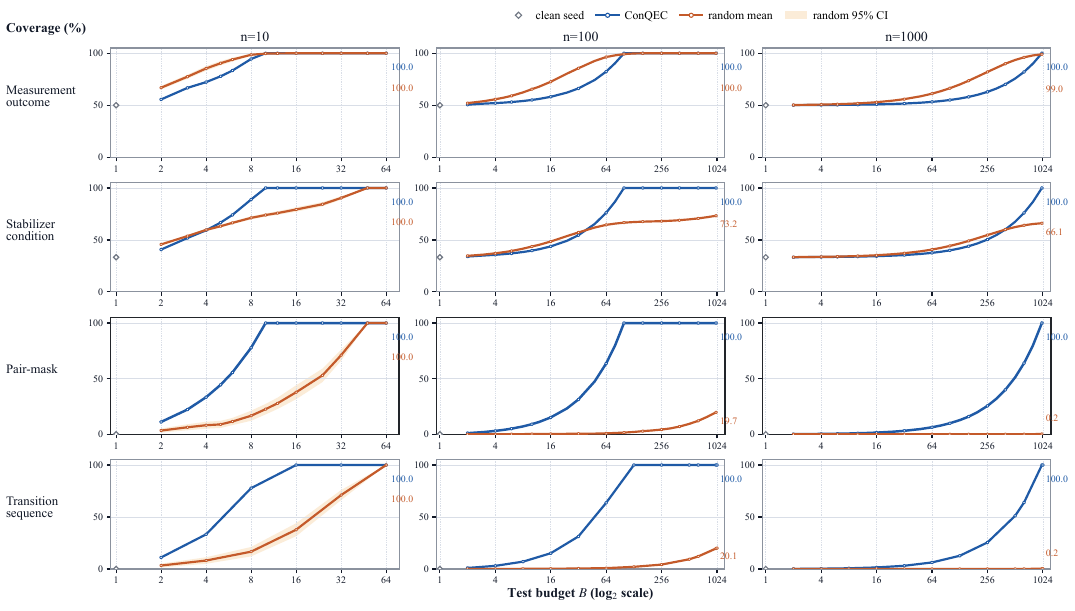}
\caption{RQ2 semantic-coverage profile on the repetition-code scale ladder.
The \(x\)-axis is the test budget \(B\) on a log\(_2\) scale, including the
clean seed. Columns vary \(n=10,100,1000\); rows report measurement-outcome,
stabilizer-condition, pair-mask, and transition-sequence coverage. Random fault
testing is averaged over 30 seeds with 95\% confidence intervals. Pair-mask
is shown separately as a representative hard sublayer of stabilizer-condition coverage.}
\label{fig:rq2-coverage-profile}
\end{figure*}

\begin{table}[h]
\centering
\caption{Pair-mask scale law over the legal weight-2 fault domain.}
\label{tab:rq2-pairmask-scale-law}
\begingroup
\scriptsize
\setlength{\tabcolsep}{2.6pt}
\renewcommand{\arraystretch}{0.96}
\begin{tabular}{@{}ccccccc@{}}
\toprule
$n$ & $|O_{\mathrm{pm}}|$ & $|\mathcal F_2|$ & $B$
& ConQEC & Random [95\% CI] & Random (exp.) \\
\midrule
10   & 9   & 45      & 5   & 44.4\% & 8.89 [5.69, 12.09]\% & 8.89\% \\
100  & 99  & 4,950   & 50  & 49.5\% & 0.61 [0.31, 0.90]\% & 0.99\% \\
1000 & 999 & 499,500 & 500 & 49.9\% & 0.10 [0.07, 0.13]\% & 0.10\% \\
\bottomrule
\end{tabular}
\endgroup

\end{table}

Figure~\ref{fig:rq2-coverage-profile} compares the methods under all three
criteria, with pair-mask reported separately as a representative hard sublayer.
On measurement-outcome coverage, random testing remains competitive: valid
weight-2 samples quickly exercise both Boolean outcomes for many exposed
checks. By the ends of the plotted campaigns, ConQEC reaches full
stabilizer-condition coverage at all three scales, while random testing falls
behind as \(n\) grows. Transition-sequence coverage
shows a similar separation at larger \(n\).

To examine the stabilizer-condition results more closely, the figure reports
pair-mask separately. Each check is counted once when paired replay confirms
faults on both support qubits (\(\mu_c(F)=2\)). The two faults cancel in parity,
giving the same check outcome as when neither support qubit is faulted;
outcome coverage cannot distinguish these cases.

To explain the scale dependence of pair-mask coverage in
Figure~\ref{fig:rq2-coverage-profile}, Table~\ref{tab:rq2-pairmask-scale-law}
relates local obligations \(|O_{\mathrm{pm}}|=n-1\) to legal fault pairs
\(|\mathcal{F}_2|=\binom{n}{2}\), where \(n\) is the number of physical
qubits. In this ladder, a legal weight-2 fault can cover at most
one pair-mask obligation, so full coverage requires at least \(n-1\)
non-clean tests, or \(B=n\) including the clean seed. We report
\(B=n/2\), half this minimum budget, to compare coverage before completion.
ConQEC covers \(44.4\%\), \(49.5\%\), and \(49.9\%\) of pair-mask
obligations, whereas random testing covers \(8.89\%\), \(0.61\%\), and \(0.10\%\) for
\(n=10,100,1000\). Each check's pair-mask activation requires one specific pair from
\(\mathcal{F}_2\). In \(s=B-1\) uniform random draws without replacement, each
such pair is selected with probability \(s/|\mathcal{F}_2|\),
which is also the expected fraction of covered pair-mask obligations.
Random (exp.) denotes this expected random coverage. At \(B=n/2\), the
budget grows linearly with \(n\) while the legal pair domain grows
quadratically, so this expectation decreases approximately as \(1/n\).

ConQEC avoids this dilution by solving directly for local objectives of the
form \(\mu_c(F)=2\), rather than waiting for uniform sampling to select both
fault atoms in a check support. Across the three criteria, ConQEC covers hard
local activations and repair-path transitions more effectively than random
testing, while random testing remains competitive on shallow outcomes.

\noindent\fbox{%
\parbox{\dimexpr\columnwidth-2\fboxsep-2\fboxrule\relax}{%
\textbf{RQ2:} Semantic guidance covers hard stabilizer obligations and
repair-path transitions more effectively than valid-input random testing,
while random remains competitive on shallow outcomes.}}

\subsection{RQ3: Scale and Cost}

RQ3 evaluates the scale boundary of ConQEC's source-level generation and the
runtime phase that limits larger campaigns. We separate three run types. A
\emph{full campaign} runs the source-level objective frontier under the campaign
budget. A \emph{single-objective run} tests whether one selected large-scale
objective can still be generated and replayed. A \emph{boundary probe} records
where a larger full campaign stops under the timeout cap.

\begin{table}[h]
\centering
\caption{RQ3 source-level scale frontier.}
\label{tab:rq3-scale-frontier}

\begingroup
\scriptsize
\setlength{\tabcolsep}{2.5pt}
\renewcommand{\arraystretch}{1.08}
\begin{tabular}{@{}>{\raggedright\arraybackslash}p{0.20\columnwidth}
                >{\raggedright\arraybackslash}p{0.33\columnwidth}
                >{\raggedright\arraybackslash}p{0.40\columnwidth}@{}}
\toprule
Family & Completed campaign & Larger run / boundary \\
\midrule
Repetition &
\(n=3000\), 3000q/2999c; 2842.8s &
Single objective: \(n=5000\), 5000q/4999c. Full campaign times out after 24 SMT queries. \\
\addlinespace[0.2em]
Surface/toric &
\(d=41\), 3362q/3362c; 957.5s &
Single objective: \(d=51\), 5202q/5202c. Full campaign times out after 5 SMT queries. \\
\addlinespace[0.2em]
Tanner/LDPC &
\(n{\sim}1000\), 1000q/1000c, \(B=128\); 105.1s &
Single objective: \(n{\sim}1000\), 1000q/1000c. Next \(n{\sim}2000\) hits a generator boundary. \\
\bottomrule
\end{tabular}
\endgroup
\end{table}

Table~\ref{tab:rq3-scale-frontier} distinguishes completed full campaigns from
larger single-objective runs. Here, \(n\) denotes the family size parameter and
\(d\) the code distance; q and c indicate physical-qubit and check counts.
Full campaigns complete through repetition \(n=3000\) and surface/toric
\(d=41\). Larger
single-objective runs still complete at repetition \(n=5000\) and surface/toric
\(d=51\), showing that individual large objectives can still be generated and
validated. However, full campaigns at those largest rows reach the timeout
boundary after writing 24 and 5 SMT queries, respectively. Thus the large-row boundary
is not concrete replay; it is repeated objective generation and solver/backend
search. Tanner/LDPC is reported only up to its supported source-level scale,
because the next requested size is a source-generator capability boundary
rather than runtime-pressure evidence.

\begin{figure}[t]
  \centering
  \includegraphics[width=\columnwidth]{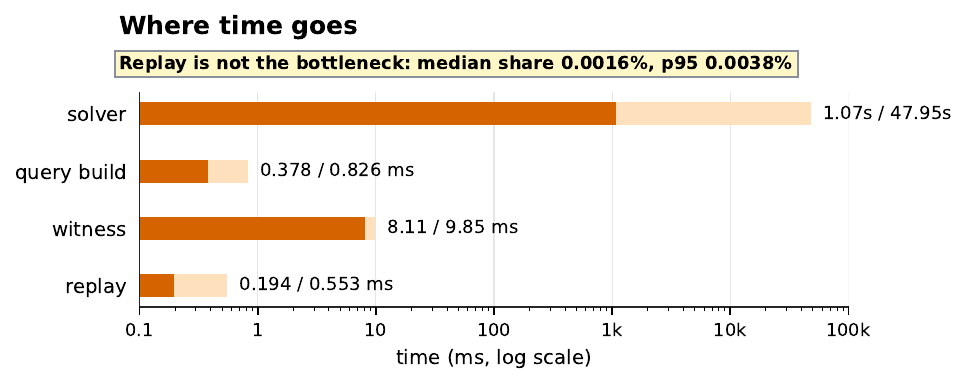}
  \caption{RQ3 phase-cost split across completed source-level runs.}
  \label{fig:rq3-replay-share}
\end{figure}

Figure~\ref{fig:rq3-replay-share} explains where time goes. Solver/backend time
dominates completed source-level runs, with median 1.07s and p95 47.95s.
Query construction and witness processing are smaller, and paired replay stays
sub-millisecond, with median 0.194ms and p95 0.553ms. Replay accounts for only
0.0016\% of wall-clock time at the median and 0.0038\% at p95. The current
scale limit of source-level generation is therefore search/query cost, not the
cost of replaying candidate faults.

\noindent\fbox{%
\parbox{\dimexpr\columnwidth-2\fboxsep-2\fboxrule\relax}{%
\textbf{RQ3:} ConQEC scales to thousand-qubit source-level campaigns; the next
scaling boundary is repeated SMT solving, while paired replay remains
negligible.}}

\subsection{RQ4: Public Artifact Applicability}
\label{sec:rq4-public-artifacts}

RQ4 asks whether the replay-backed evidence contract used on generated
source-level subjects can also be instantiated on public QECC artifacts. We use
the two public-artifact settings introduced in the experimental setup: a
Tesseract detector-adapter setting and a PyMatching wrapper-regression setting.
Together, they exercise two different public-artifact interfaces: detector-level
predicates exported from a public Stim corpus, and wrapper-level decoder
contracts for parent/fixed regression replay.

\begin{table}[t]
\centering
\caption{RQ4 public-artifact applicability.}
\label{tab:rq4-public-artifact-results}
\begingroup
\footnotesize
\setlength{\tabcolsep}{3.0pt}
\renewcommand{\arraystretch}{1.10}
\captionsetup[subtable]{position=top,justification=raggedright,singlelinecheck=false,labelfont=bf,textfont=bf,farskip=0pt,nearskip=3pt}
\subfloat[Tesseract detector adapter\label{tab:rq4-tesseract}]{%
\begin{tabular}{@{}>{\raggedright\arraybackslash}m{0.34\columnwidth}*{4}{>{\centering\arraybackslash}m{\dimexpr(0.66\columnwidth-8\tabcolsep)/4\relax}}@{}}
\toprule
\multirow{2}{*}{Comparison outcome} & \multirow{2}{*}{Cases} & \multicolumn{3}{c@{}}{Diagnostic records} \\
\cmidrule(l){3-5}
& & Decoder & Logical & Total \\
\midrule
Identical outputs & 40,436 & 0 & 0 & 0 \\
Order-only difference & 326 & 326 & 0 & 326 \\
Logical disagreement & 36 & 36 & 36 & 72 \\
\midrule
Total & 40,798 & 362 & 36 & 398 \\
\bottomrule
\end{tabular}%
}\par\medskip
\subfloat[PyMatching previously fixed regression\label{tab:rq4-pymatching}]{%
\begin{tabular}{@{}>{\centering\arraybackslash}m{0.08\columnwidth}>{\centering\arraybackslash}m{0.20\columnwidth}*{2}{>{\raggedright\arraybackslash}m{\dimexpr(0.72\columnwidth-6\tabcolsep)/2\relax}}@{}}
\toprule
Test & Syndrome & Parent replay & Fixed replay \\
\midrule
1 & \texttt{00000} & Correction \texttt{0000} & Correction \texttt{0000} \\
2 & \texttt{10000} & Correction \texttt{1111} & Correction \texttt{1111} \\
3 & \texttt{01000} & Correction \texttt{0000} & Correction \texttt{0000} \\
4 & \texttt{00100} & Correction \texttt{0000} & Correction \texttt{0000} \\
5 & \texttt{00010} & No perfect matching & Correction \texttt{0000} \\
\bottomrule
\end{tabular}%
}
\endgroup
\end{table}

The Tesseract subset is fixed to 173 non-\texttt{REPEAT} subjects before
inspecting outcomes. Exported XOR predicates are validated against the
corresponding detector error model (DEM) rows with zero mismatches, then used
for generation and paired replay. We compare the Tesseract and Simplex decoders
from the same library on the corresponding DEMs, using 173 clean and
40,625 single-fault inputs. For 40,436 inputs, both decoders return identical
error lists and logical predictions, producing no diagnostic records. The
remaining 362 fault cases yield 398 diagnostic records
(Table~\ref{tab:rq4-public-artifact-results}): 326 return the
same DEM error set in different orders with identical logical prediction; 36
disagree on logical-observable support, each yielding decoder- and
recovery/logical-level records (\(398=326+2\times36\)). Order-only records arise
from our order-sensitive comparison.

For PyMatching, ConQEC receives the public graph and boundary-node contract but
no failing-syndrome hint. It generates candidate syndromes from public
decoder-contract objectives and replays each generated syndrome on the
historical parent and fixed commits for a documented, previously fixed
regression. In Table~\ref{tab:rq4-public-artifact-results}(b), \emph{Correction}
is the decoder's four-bit output for the graph edges: \texttt{0000} selects none
and \texttt{1111} selects all four. The generic no-hint trace
reaches the decoder-boundary regression at the fifth generated test, when the parent fails
and the fixed version returns a correction. The parent/fixed difference is
exposed by 24 of the 32 candidate syndromes.

\noindent\fbox{%
\parbox{\dimexpr\columnwidth-2\fboxsep-2\fboxrule\relax}{%
\textbf{RQ4:} ConQEC can instantiate replay-backed FSDR evidence on public
QECC artifacts through detector-adapter execution and wrapper-level
decoder-regression replay.}}

\section{Discussion and Threats to Validity}
\label{sec:discussion}
\textbf{Scope of claims.}
ConQEC credits coverage and reports bugs only after paired actual/reference
replay records the relevant repair-path evidence. Source-level QECCs use the
QSE-backed symbolic-stabilizer interface, Tesseract uses a detector adapter, and
PyMatching uses a wrapper-level decoder contract. Results based on finite
evaluators or replay-only adapters are labeled as such and are not claimed as
source-level symbolic generation.

\textbf{Construct validity.}
RQ1 is a seeded localization study, so it measures whether exposed repair-path
bugs can be generated and localized as FSDR witnesses, not their natural
prevalence. RQ2 evaluates semantic coverage on repetition codes with legal
weight-2 faults against uniform sampling; other code families or fault models
may stress different layers. RQ1 provides the separate fault-localization evidence.
RQ3 reports empirical scale under our backend and timeout
configuration, not an asymptotic bound. RQ4 checks whether the same evidence
contract can be instantiated on public artifacts. The 36 Tesseract logical
disagreements establish behavioral differences; assigning fault to either
decoder requires further validation.

\textbf{External validity.}
ConQEC relies on the semantics exposed by each subject or adapter. Transition
objectives require explicit decoder and recovery relations; otherwise those
fields remain replay-observed evidence rather than generation targets. New
interfaces require reference relations from the code or protocol specification
and manual replay mappings. Broader
Tesseract coverage would require structured handling of \texttt{REPEAT} blocks,
and temporal-history bugs would require history predicates. More decoder
versions, corpus subsets, and richer fault histories remain future work.

\textbf{Reproducibility.}
For each reported experiment, raw logs and table-derivation manifests record
the subject, adapter or method, seed and budget when applicable, backend and
replay status, covered obligations, and FSDR witness fields where applicable. These records
regenerate the tables and make each row checkable against its stated evidence
boundary.

\section{Related Work}
\label{sec:related}
\textbf{Quantum software testing and debugging.}
Quantum software-engineering research has studied coverage
criteria~\cite{ali2021coverage}, coverage-guided
generation~\cite{wang2021quito,ye2023quratest}, combinatorial and
search-based testing~\cite{wang2022qusbt,qucat2023tool}, mutation
analysis~\cite{mendiluze2021muskit,fortunato2022mutation},
metamorphic testing~\cite{paltenghi2023morphq}, bug
benchmarks~\cite{paltenghi2022bugs,zhao2023bugs4q}, static
analysis~\cite{qchecker2023arxiv}, multi-subroutine
testing~\cite{long2024multi}, runtime
assertions~\cite{huang2019statistical}, compiler
verification~\cite{li2020projection,liu2020dynamic}, and program
reasoning~\cite{tao2022giallar}. These techniques establish important
testing and analysis foundations for general quantum programs. ConQEC targets a
different software problem: QECC implementations are repair pipelines whose
inputs are Pauli faults and whose useful debugging evidence is not just a final
failure, but the fault, syndrome, decoder decision, recovery action, oracle
result, and first divergent stage.

\textbf{QEC infrastructure and verification.}
Quantum error correction (QEC) engineering provides mature infrastructure for
simulation, decoding, and
public artifacts, including Stim~\cite{gidney2021stim},
qecsim~\cite{tuckett2020qecsim}, PyMatching~\cite{higgott2022pymatching}, and
Tesseract~\cite{beni2025tesseract}. These systems make large QEC workflows
available to tools, but they do not by themselves define a concolic testing
workflow with semantic coverage and replay-backed bug witnesses. Formal QEC
methods such as QSE~\cite{fang2024qse}, QECV~\cite{wu2021qecv},
Veri-QEC~\cite{huang2025veriqec}, and fault-tolerance
verification~\cite{chen2025faulttolerance} show how to symbolically execute,
check, or verify QEC/QECC programs against formal obligations. ConQEC is
complementary to these verification tools: it uses symbolic or table-backed QEC
reasoning to generate concrete Pauli faults under a test budget, then relies on
actual/reference replay to decide coverage and accept developer-facing FSDR
witnesses.

\textbf{Quantum concolic testing.}
Quantum Concolic Testing (QCT)~\cite{xia2025qct} is the closest prior
concolic-testing framework for general quantum programs. It is valuable because
it brings concolic testing to quantum software, but its constraints are
amplitude-based: reasoning about an \(n\)-qubit state requires tracking a
\(2^n\)-dimensional state space, and its reported branch-coverage evaluation
focuses on programs with fewer than five qubits. ConQEC instead specializes
concolic testing to QECC repair paths. It represents Pauli faults, syndrome and
detector predicates, decoder choices, recovery actions, and logical checks as
compact repair-path constraints. This lets ConQEC scale to QECC families with
up to thousands of qubits while returning replay-valid FSDR witnesses that
localize the software stage where the implementation diverges from the
reference.

\section{Conclusion}
\label{sec:conclusion}

This paper presented ConQEC, a concolic testing framework that treats QECC
execution as a replay-backed repair path. ConQEC avoids amplitude-level quantum
state reasoning by generating concrete Pauli faults from compact repair-path
constraints over syndromes, decoder decisions, recovery actions, and logical
checks. In our evaluation on generated source-level QECC families and public
QECC artifacts with up to thousands of qubits, ConQEC detects the seeded bug in
every experimental case and reports a replay-valid, fault-localizing FSDR
witness for each. The remaining results show that semantic objectives improve
hard repair-path coverage, paired replay is not the scale bottleneck, and the
same evidence contract can be instantiated on public QECC artifacts.

\section*{Data Availability}
The ConQEC source code and experimental artifacts supporting this study are available at
\url{https://github.com/Hikari-HN/conqec-qse-artifact}.

\bibliographystyle{IEEEtran}
\bibliography{refs}

\end{document}